\documentclass[prb,showpacs,twocolumn,preprintnumbers,amsmath,amssymb,floatfix]{revtex4}
\usepackage[dvips]{graphicx}
\usepackage[latin1]{inputenc}
\begin{document}
\draft

\newcommand{\lxpc} {Li$_{x}$ZnPc }
\newcommand{\lpc} {Li$_{0.5}$MnPc }
\newcommand{\etal} {{\it et al.} }
\newcommand{\ie} {{\it i.e.} }
\newcommand{\ip}{${\cal A}^2$ }

\hyphenation{a-long}

\title{Low-energy spin dynamics in the [YPc$_2$]$^0$ $S=1/2$ antiferromagnetic chain}

\author{F. Branzoli$^{1}$, P. Carretta$^{1}$, M. Filibian$^{1}$, S. Klyatskaya$^{2}$ and M. Ruben$^{2}$}

\address{$^{1}$Department of Physics ``A. Volta'', University of Pavia-CNISM, 27100 Pavia (Italy)}
\address{$^{2}$Institute of Nanotechnology, Karlsrhue Institute of Technology (KIT), 76344 Eggenstein-Leopoldshafen (Germany)}

\widetext

\begin{abstract}

$^1$H nuclear magnetic resonance (NMR) measurements in [YPc$_2$]$^0$, an organic compound formed by radicals
stacking along chains, are presented. The temperature dependence of the macroscopic susceptibility, of the NMR
shift and of the spin-lattice relaxation rate $1/T_1$ indicate that the unpaired electron spins are not
delocalized but rather form a $S=1/2$ antiferromagnetic chain. The exchange couplings estimated from those
measurements are all in quantitative agreement. The low-energy spin dynamics can be described in terms of
diffusive processes and the temperature dependence of the corresponding diffusion constant suggests that a
spin-gap around 1 K might be present in this compound.

\end{abstract}

\maketitle

\narrowtext

\section{Introduction}

Molecular solids have attracted much interest since decades owing to the possibility to easily tune their
properties either through a chemical bottom up approach or by varying physical parameters as the external
pressure.\cite{RevMol} One of the most versatile families of molecular solids is the one based on phthalocyanine
(Pc= C$_{32}$H$_{16}$N$_8$) molecules.\cite{Kaldis} In fact, the employment of these materials in different areas,
ranging from the fabrication of organic light emitting diodes, to contrast agents or spintronics materials, has
been envisaged. Pc-based compounds have attracted significant interest in the last decade after it has been
suggested that high temperature superconductivity could be induced in these materials by alkali
doping\cite{Tosatti} and, more recently, when it has been recognized that neutral [LnPc$_2]^0$ molecules, with Ln
a lanthanide ion, are molecular nanomagnets with extremely long coherence times at liquid nitrogen
temperature.\cite{FraRap,FraJACS,Ishi1,Ishi2,FraPRB} Owing to the flat shape of Pc molecules, the structure of
Pc-based materials is typically characterized by chains along which Pc molecules tend to stack.\cite{StrucPc}
Accordingly some of the Pc-based materials show many similarities to the Beechgaard salts.\cite{Beech}

Bis(phthalocyaninato) yttrium [YPc$_2]^0$ compound can be considered the parent compound of the aforementioned
[LnPc$_2]^0$ molecular magnets. In fact, it is characterized by the absence of localized $f$ electrons and the
microscopic properties are mainly associated with the presence of an unpaired electron delocalized in the $a_2$
$\pi$ orbital, due to the one-electron oxidation of the [YPc$_2]^-$ unit.\cite{Elec} Thus, [YPc$_2]^0$ allows to
investigate the spin dynamics associated only with this unpaired electron spin, independently from the one due to
$f$ electrons. One of the most suitable tools to address this aspect is nuclear magnetic resonance (NMR)
technique. In this work we present an experimental study of the magnetic properties of [YPc$_2]^0$ compound by
means of magnetization and nuclear magnetic resonance (NMR) measurements. The temperature dependence of the
macroscopic susceptibility, of the NMR shift and of the spin-lattice relaxation rate $1/T_1$ clearly show that
this system is a prototype of a $S=1/2$ antiferromagnetic chain, characterized by a diffusive low-frequency spin
dynamics and, possibly, by the presence of a low-energy spin-gap.

\begin{figure}[h!]
\vspace{6.5cm} \includegraphics{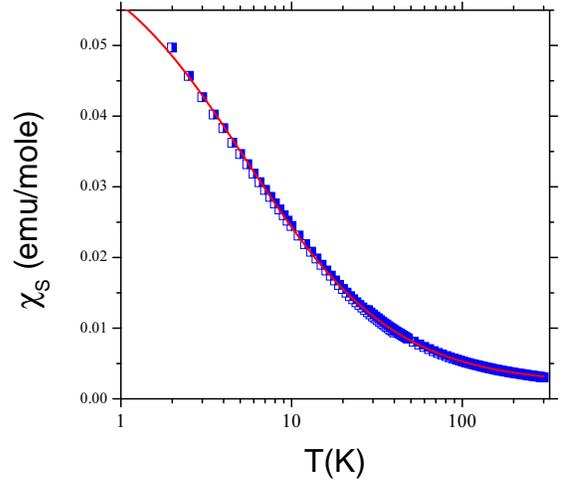} \caption{\label{chiY}
Temperature dependence of static uniform susceptibility  $\chi_S$ for [YPc$_2]^0$ complex, derived from SQUID
magnetization measurements. The solid line shows the best fit of the data to Curie-Weiss law.}
\end{figure}

\section{Experimental Results and Discussion}

[YPc$_2]^0$ polycristalline samples were synthesized by using some modifications of the protocol published in Ref.
\onlinecite{Moussavi}. All reagents were purchased from Across or Aldrich and used without further purification. A
mixture of 1,2-dicyanobenzene (42 mmol), Y(acac)$_3$ ·4H$_2$O (5 mmol), and 1,8-diazabicyclo[5,4,0]undec-7-ene
(DBU) (21 mmol) in 50 mL of 1-pentanol was refluxed for 1.5 days. The solution was allowed to cool to room
temperature. The precipitate was collected by filtration and washed with $n$-hexane and Et$_2$O. The crude purple
product was redissolved in 800 ml of CHCl$_3$/MeOH (1/1) and undissolved PcH$_2$ was filtered off. Both forms,
blue (anionic [YPc$_2]^-$) and green (neutral [YPc$_2]^0$), were obtained simultaneously, as revealed by
electronic absorption spectra. In order to convert the unstabilized anionic form to the neutral one, the reaction
mixture was presorbed on active (H$_2$O-$0\%$) basic alumina oxide. Purification was carried out by column
chromatography on basic alumina oxide (deactivated with 4.6\% H$_2$O, level IV) with chloroform methanol mixture
(10:1) as eluent. In general, the yield was 30-35\%. According to microelemental analysis based on atomic
spectroscopic methods (ICP) performed at \textit{Mikroanalytisches Labor Pascher}, the powder sample contains
molecules of [YPc$_2]^0$, water and CH$_2$Cl$_2$ in ratio 1:1:1/3. The molecules crystallized in the space group
P$2_12_12_1$ ($\gamma$-phase), as reported in Ref. \onlinecite{Struc}.

DC magnetization ($M$) measurements have been performed by using an MPMS-XL7 Quantum Design superconducting
quantum interference device (SQUID) magnetometer. The magnetization was found to depend linearly on the magnetic
field intensity $H$, for $H\leq 5$ kGauss, over all the explored temperature range and, accordingly, the
macroscopic static uniform susceptibility can be written as $\chi_S= M/H$. The temperature dependence of $\chi_S$
reveals the presence of antiferromagnetic correlations. In fact, $\chi_S$(T) can be nicely reproduced by a
Curie-Weiss (CW) law
\begin{equation}
 \label{chi}
  \chi_S(T) = {C \over T + \Theta} + \chi_0
  \;\;\;\; ,
\end{equation}
where $C = g^2\mu_B^2S(S +1)N_A/(3k_B)$ is Curie constant ($\mu_B$ the Bohr magneton, $g$ the Land\'e factor,
$N_A$ Avogadro's number and $k_B$ Boltzmann constant). $\Theta$ is the CW temperature and $\chi_0$ a temperature
independent term mainly due to diamagnetic and Van-Vleck corrections. By fitting the data, leaving all three
parameters free, we found an antiferromagnetic CW temperature $\Theta = 5.37±0.04$ K (Fig. \ref{chiY}) and a Curie
constant $C= 0.342\pm 0.002 $ erg $\cdot$ K/G$^2$, quite close to the value 0.375 erg$\cdot$K/G$^2$, expected for
a $S=1/2$ system. If we fixed $C=0.375$ erg$\cdot$K/G$^2$ the fit was still good and the CW temperature $\Theta=
6.18\pm 0.03$ K. The temperature dependence of $\chi_S$ shows that the unpaired electron spins are localized along
the chains formed by [YPc$_2$]$^0$ molecules, although a certain overlap of the $\pi$ orbitals of adjacent
molecules must be present in order to justify the magnitude of the antiferromagnetic exchange coupling $J_e$. In
fact, although this system should present a narrow half-filled band, the sizeable Hubbard Coulomb repulsion $U\sim
1 eV$ prevents the electron delocalization along the chain.\cite{Hubb} In this limit, $J_e= \Theta= 4t^2/U$, with
$t$ the hopping integral among adjacent molecules. From the estimated value of $\Theta$ one would derive a band
width formed by the overlap of $a_2$ orbitals in adjacent molecules $W= 4t\sim 0.05$ eV$\ll U$, justifying the
spin localization along the chain.

\begin{figure}[h!]
\vspace{7cm} \includegraphics{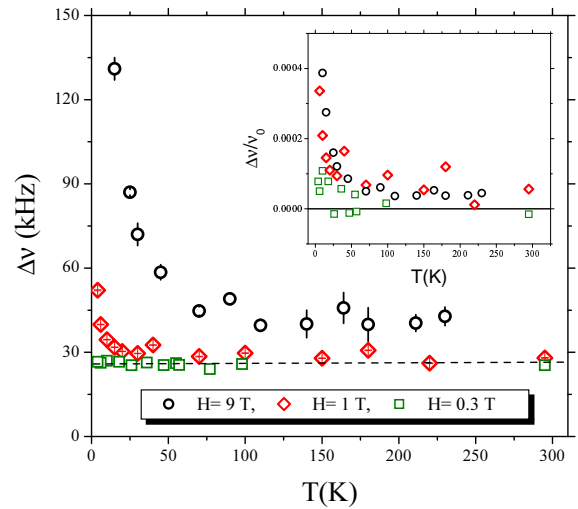} \caption{\label{Dnu}
Temperature dependence of $^1$H full NMR linewidth at half intensity in [YPc$_2]^0$, at three different magnetic
fields. In the inset the linewidth is normalized by the Larmor frequency after subtracting the constant linewidth
due to nuclear dipole-dipole interaction.}
\end{figure}

The $^1$H NMR spectra were obtained in the 1.6-300 K temperature range for magnetic field intensities $H = 9$ T, 1
T and 0.3 T. The spectra were derived from the Fourier transform of half of the echo formed after a $\pi/2 -\tau
-\pi$ pulse sequence, when the full NMR line could be irradiated or, otherwise, from the envelope of the echo
amplitude upon varying the irradiation frequency. The line-shape was gaussian in all the investigated temperature
range. For $H = 9$ T and $H = 1$ T a broadening of the spectrum can be observed at low temperature, which is more
pronounced at higher field intensities (Fig. \ref{Dnu}). On the other hand, for $H = 0.3$ T the linewidth
$\Delta\nu$ is practically temperature independent and the broadening is likely to be due just to nuclear
dipole-dipole interaction. The increase of the linewidth with $H$ suggests that the low temperature line
broadening originates from some anisotropy in the hyperfine coupling, which for a powder gives rise to a linewidth
proportional to the susceptibility. In fact, it is noticed that if we subtract the $T$-independent contribution at
$H= 0.3$ Tesla from the raw data and divide the linewidth by the Larmor frequency $\nu_0$, the data at different
fields overlap (inset to Fig.\ref{Dnu}). This result also indicates that there is not an additional internal field
due to the onset of a long-range magnetic order down to 1.6 K.

\begin{figure}[h!]
\vspace{7cm} \includegraphics{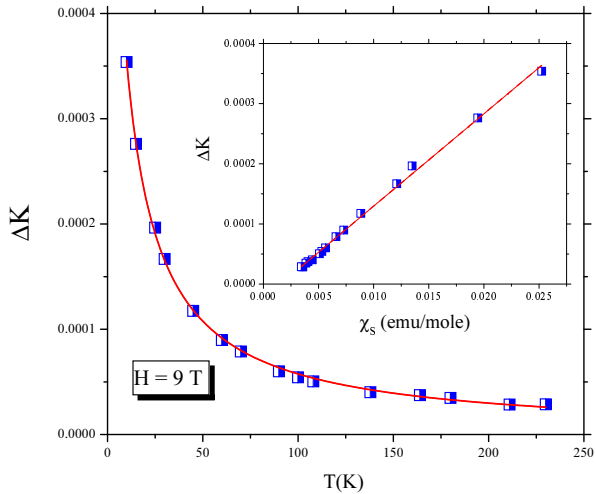} \caption{\label{line-K}
Temperature dependence of $^1$H paramagnetic shift $\Delta K$ in [YPc$_2]^0$.  The solid line shows the best fit
of the data to Curie-Weiss law. In the inset $\Delta K$ is reported as a function of the macroscopic
susceptibility. The solid line is the best fit according for a linear dependence of $\Delta K$ vs $\chi_S$. }
\end{figure}

The NMR paramagnetic shift $\Delta K = (\nu_R - \nu_0)/\nu_0$, with $\nu_R$ the resonance frequency, shows a more
pronounced temperature dependence (Fig. \ref{line-K}). As expected, it was found to increase upon cooling,
according to
\begin{equation}
 \label{KvsChi}
  \Delta K= {A \chi_S \over 2\mu_B N_A}
  \;\;\;\; ,
\end{equation}

namely the temperature dependence of $\Delta K$ should be the same of the macroscopic spin susceptibility. In
fact, also $\Delta K(T)$ is found to obey a Curie-Weiss law with a Curie-Weiss temperature $\Theta = 7.4 K \pm
0.3$ K, close to the one derived from SQUID magnetization measurements. The small difference between those two
type of measurements could be associated with a tiny amount of impurities which might affect the macroscopic
susceptibility. Accordingly, the measurement of the microscopic susceptibility with paramagnetic shift
measurements is expected to provide a more reliable estimate of the static uniform spin susceptibility and of the
Curie-Weiss temperature. By plotting $\Delta K$ as a function of $\chi_S$ a linear trend is attained (Fig.
\ref{line-K}) and from the slope it is possible to estimate the isotropic term of the hyperfine coupling tensor
$A=180\pm 10$ Gauss.

\begin{figure}[h!]
\vspace{7cm} \includegraphics{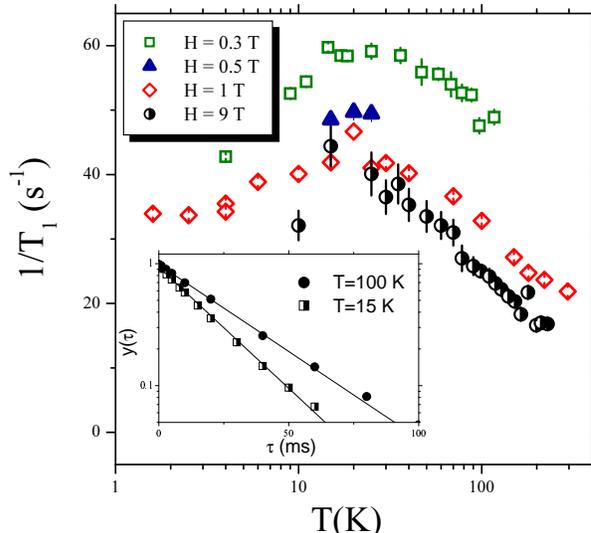} \caption{\label{T1vsT} $^1$H
nuclear spin-lattice relaxation rate temperature dependence for [YPc$_2]^0$ compound measured for different values
of the applied field. In the inset the recovery of the nuclear magnetization as a function of the delay $\tau$
between the saturating and the echo readout sequences is shown at two different temperatures. The solid lines show
the best fit for a single exponential recovery.}
\end{figure}

The $^1$H nuclear spin-lattice relaxation rate $1/T_1$ was measured in the 1.6 - 300 K temperature range and for
different values of the external field. 1/T$_1$ was extracted from the recovery of nuclear magnetization after a
saturation recovery pulse sequence. The recovery law was found to be a single exponential in all the explored
temperature range (Fig. \ref{T1vsT}, inset). This result is an evidence that the unpaired electron is delocalized
onto a $\pi$ orbital within the molecule. In fact, since in the two phthalocyanine rings a large number of
inequivalent proton sites is present, if the electron was on a more localized orbital a distribution of hyperfine
couplings would be present and, accordingly, a stretched exponential recovery law should be observed. Moreover,
the fact that hyperfine coupling seems quite isotropic indicates that it could originate from the contact
interaction between the unpaired electron spin in the $a_2$ $\pi$ orbital and the $^1$H nuclei.

\begin{figure}[h!]
\vspace{6.5cm} \includegraphics{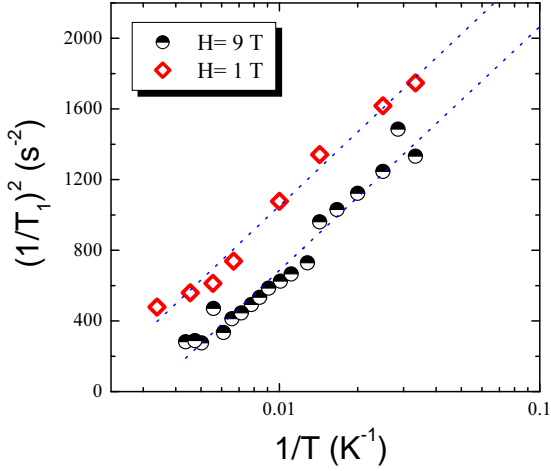}
\caption{\label{Ysquared} The 1/T$_1$ squared is plotted as a function of T$^{-1}$, in logarithmic scale, for two
values of the external field (H = 9 T, circles and H = 1 T, squares). The dashed lines represent the best fits to
Eq. (\ref{log}).}
\end{figure}

The temperature dependence of $1/T_1$ at different magnetic fields is shown in Fig. (\ref{T1vsT}). In general, for
a relaxation process driven by electron spin fluctuations one can write
\begin{equation}
\frac{1}{T_1}= \frac{\gamma^2}{2N}\sum_{\alpha,\mathbf{q}} (|A_{\mathbf{q}}|^2
S_{\alpha,\alpha}(\mathbf{q},\omega_L))_{\perp}
  \;\;\;,
\end{equation}
where $\gamma$ is the nuclear gyromagnetic ratio, $|A_{\mathbf{q}}|^2$ the form factor describing the hyperfine
coupling with spin excitations at wave-vector $\mathbf{q}$ and $S_{\alpha,\alpha}(\mathbf{q},\omega_L)$
($\alpha=x,y,z$) the component of the dynamical structure factor at the Larmor frequency. In the high temperature
limit, namely when the thermal energy is much larger than the exchange energy ($T \gg \Theta$), the 1/T$_1$ of a
spin $S=1/2$ antiferromagnet becomes temperature and field independent and is given by \cite{Moriya}
\begin{equation}
 \label{1suT1par}
    {1 \over T_1} = {\gamma^2 \over 2} (A_x^2+A_y^2){S(S+1) \over 3}{{\sqrt 2 \pi} \over \omega_H}
 \;\;\; ,
\end{equation}
where $A_x\simeq A_y\simeq A$ are the components of the hyperfine
coupling tensor which is basically isotropic, while $\omega_H=
(J_ek_B/\hbar) \sqrt{2zS(S + 1)/3}$ is the Heisenberg exchange
frequency, with $z= 2$ the number of nearest neighbour spins along
the chain. By taking the measured value of $1/T_1\simeq 20$
s$^{-1}$ at high temperature, from Eq. (\ref{1suT1par}) it is
possible to estimate the exchange frequency $\omega_{H}\simeq
9.2\cdot 10^{11}$ rad/s, corresponding to an exchange coupling
constant $J_e\simeq$ 7.0 K, in quite good agreement with the value
which can be estimated from the NMR shift measurements. Upon
decreasing the temperature, for $200$ K$\geq T\geq 30$ K, one
observes a progressive slow increase of $1/T_1$ (Fig.
(\ref{T1vsT})). In particular, it is noticed that nuclear
spin-lattice relaxation rate increases on decreasing temperature
according to
\begin{equation}
\label{log}
 1/T_1 \propto ln^{1/2} (T_0/T)
       \;\;\; .
\end{equation}
In fact, in Fig. (\ref{Ysquared}) one observes that $(1/T_1)^2$ is
a linear function of $1/T$, when reported in logarithmic scale.
Remarkably, this logarithmic increase of $1/T_1$ is expected  in a
$S=1/2$ Heisenberg antiferromagnet, but for $T\leq J_e$
\cite{Takigawa}. Here it is not clear why the logarithmic behavior
extends up to $T\gg J_e$.

At about 20 K a peak in the nuclear spin-lattice relaxation rate
appears (Fig. \ref{T1vsT}), whose intensity decreases by
increasing the external field intensity. Eventually, below
$T\simeq 5$ K, the 1/T$_1$ is only weakly temperature dependent.
The maximum in $1/T_1$, not associated with molecular motions,
could be due to a form factor, which partially filters out the
antiferromagnetic fluctuations as the system gets more and more
correlated.

\begin{figure}[h!]
\vspace{7cm} \includegraphics{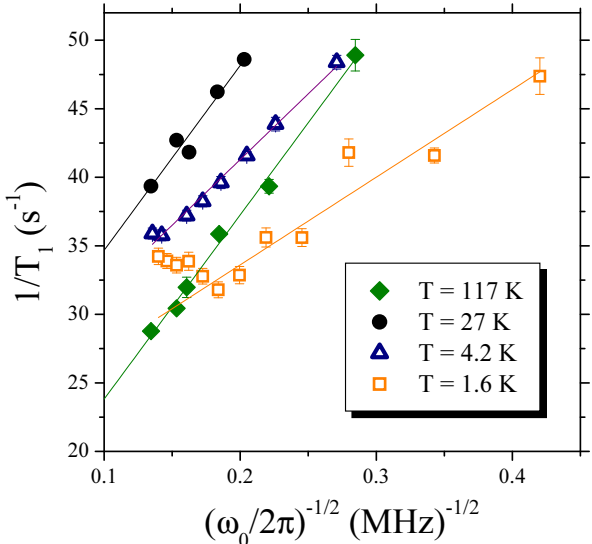} \caption{\label{T1vsH} The
$^1$H spin-lattice relaxation time 1/T$_1$ in [YPc$_2]^0$ is plotted as a function of $(\omega_0/2\pi)^{-1/2}$ for
different selected temperatures. The solid lines show the best fit according to Eqs.\ref{1suT1diff} and
\ref{Jdiff} in the text.}
\end{figure}

The magnetic field dependence of $1/T_1$ (Fig. \ref{T1vsH}) can originate from the diffusive nature of the spin
correlation function, which in one dimension is characterized by long-time tails yielding to a divergence of the
low-frequency spectral density $J(\omega)$.\cite{Boucher} In fact, in the presence of diffusive processes for the
spin excitations $1/T_1$ can be written in terms of the spectral density for the spin excitations according to the
following equation \cite{Devreux}:
\begin{equation}
 \label{1suT1diff}
    {1 \over T_1} = {\gamma^2 \over 2} {k_BT\chi_0 \over (g\mu_B)^2}
    \left [{3 \over 5} A_d^2J(\omega_0)+\left (A^2+{7 \over 5} A_d^2\right )J(\omega_e\pm \omega_0) \right ]
 \;\;\;,
\end{equation}
\begin{figure}[h!]
\vspace{6.5cm} \includegraphics{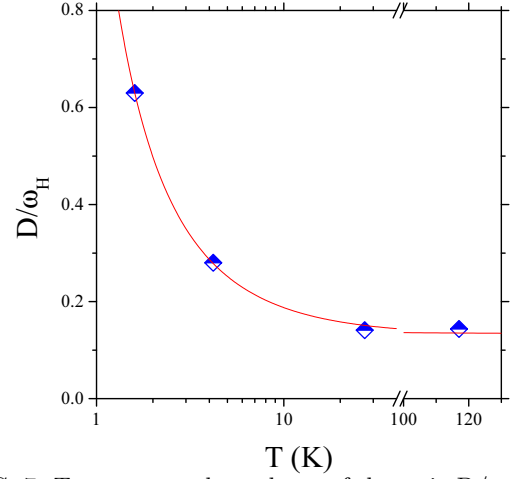} \caption{\label{DvsT}
Temperature dependence of the ratio D/$\omega_{H}$ between the spin diffusion coefficient $D$ and the exchange
frequency $\omega_{H}$ in [YPc$_2]^0$ compound as derived from the slopes in the 1/T$_1$ vs
$(\omega_0/2\pi)^{-1/2}$ plots in Fig. \ref{T1vsH}. The solid line gives the best fit according to $D\propto
exp(\Delta/T)$ with $\Delta=1.2\pm 0.4$ K.}
\end{figure}
where $A_d$ is the anisotropic term of the hyperfine coupling, which hereafter shall be neglected since $A^2\gg
A_d^2$. Then just the second term in square bracket can be considered. In Eq.\ref{1suT1diff} $\chi_0$ is the
static uniform susceptibility per spin and $\omega_e = \omega_0\gamma_e/\gamma$ is the electron resonance
frequency. This means that during the nuclear relaxation process a simultaneous flip of the electron and nuclear
spins occur, involving an energy exchange $\hbar(\omega_e \pm \omega_0)$, and 1/T$_1$ thus probes the spin
excitations at a frequency close to $\omega_e$.

In a one dimensional system, the spectral density at $\omega_e$ is characterized by a low-frequency divergence
given by \cite{Soda}
\begin{equation}
 \label{Jdiff}
    J(\omega_e) = {1 \over \sqrt {2D}} ({\omega_c+{\sqrt {\omega_e^2+\omega_c^2}} \over \omega_e^2+\omega_c^2})^{1/2}
 \;\;\;,
\end{equation}
where $\omega_c$ is a low-frequency cutoff accounting for the
finite spin anisotropy and/or inter-chain coupling, while $D$ is
the spin diffusion rate. In Fig. (\ref{T1vsH}), the 1/T$_1$ is
plotted as a function of $\nu_0^{-1/2}$. The observed linear trend
further proves the one-dimensional nature of the antiferromagnetic
correlations. Moreover, the absence of a low-frequency flattening
in 1/T$_1$ plot indicates that spin diffusion occurs in the
electronic frequency range $\omega_c\ll \omega_e\ll D$. Thus, from
the slopes of the curves it is possible to deduce the spin
diffusion coefficient at different temperatures (Fig. \ref{DvsT})
considering $A\simeq$ 180 Gauss and neglecting $\omega_c \ll
\omega_e$ in Eqs (\ref{1suT1diff}-\ref{Jdiff}). The estimated spin
diffusion coefficient is of the order of the exchange frequency
$\omega_{H}$ and it is found to progressively decrease with
temperature and to become nearly constant above 20 K. It is
interesting to observe that $D\propto exp(\Delta/T)$, namely the
behaviour expected for one-dimensional antiferromagnets in the
presence of a spin-gap $\Delta$ between singlet and triplet
excitations.\cite{Damle} Here we find that $\Delta=1.2\pm 0.4$ K
suggesting that a small gap, either due to competing exchange
interactions or to a dimerization might be present in [YPc$_2]^0$.
It is interesting to observe that, at low temperature, when the
Zeeman energy $\hbar\omega_e\simeq \Delta$ the breakdown of
Eq.\ref{Jdiff} is noticed. In fact, in Fig. (\ref{T1vsH}) one
clearly notices that at $T= 1.6$ K the linear behaviour is no
longer obeyed at high fields (i.e. low values for
$\sqrt{2\pi/\omega_0}$) and $1/T_1$ ceases to decrease with
increasing field. This could be due to the modifications in the
spin correlations induced by the magnetic field for
$\hbar\omega_e\simeq \Delta$, possibly associated with the
progressive closure of the spin gap.


In conclusion, from magnetization, $^1$H NMR paramagnetic shift and $T_1$ measurements we have derived the
magnitude of the antiferromagnetic exchange interaction in [YPc$_2]^0$ compound and found an overall good
agreement. The low-energy spin excitations are of diffusive character and characteristic of one-dimensional
antiferromagnets. From the temperature dependence of the spin diffusion rate derived from $1/T_1$ vs. $H$
measurements it was found that a spin-gap around 1 K might be present in this compound.

\section*{Acknowledgements}

The research activity in Pavia was supported by Fondazione Cariplo (Grant N. 2008-2229) research funds.



\end{document}